\documentclass[aps, prl, reprint]{revtex4-2}

\usepackage{graphicx}
\usepackage{xcolor}
\usepackage{amsmath}
\usepackage{hyperref}

\begin{document}

\title{Interface Fragmentation via Horizontal Vibration: A Pathway to Scalable Monodisperse Emulsification}

\author{Linfeng Piao}
\author{Anne Juel}%
 \email{anne.juel@manchester.ac.uk}
\affiliation{%
 \textit{Department of Physics and Astronomy and Manchester Centre for Nonlinear Dynamics, The University of Manchester, Oxford Road, Manchester M13 9PL, United Kingdom} }  

\begin{abstract}
We present a scalable method for producing monodisperse micro-scale emulsions in a rectangular container holding two stably stratified layers of immiscible liquids by applying horizontal vibration. \textcolor{black}{This setup enables the excitation of a single line of ordered Faraday waves along each end wall when viscous forces dominate interfacial dynamics. Our experiments and theoretical modeling show that the critical non-dimensional acceleration for the breakup of the wave tips in a regular array of droplets scales as}
$N^{-1/2} \omega^{*3/2}$, where $N$ is the \textcolor{black}{kinematic} viscosity ratio and $\omega^{*}$ is the frequency of forcing on the viscous-capillary scale. The droplet diameter can be easily tuned by varying the forcing parameters, \textcolor{black}{and the number of droplets generated per cycle is proportional to the width of the container.}
\end{abstract}

\keywords{Suggested keywords}
\maketitle

Emulsification, a powerful technique for dispersing a liquid in a second immiscible phase, underpins a wide range of practical applications from food processing \cite{McClements2004} and drug delivery systems \cite{Washington1996,Ho2022} to solvent extraction \cite{Ahmad2011}. Although the processes used in these applications are different, they typically demand precise control of the emulsion properties. For example, droplet size, \textcolor{black}{stability}, and production rate are key determinants of the mass transfer rate of the solvent to be extracted using emulsion-liquid-membrane (ELM) \textcolor{black}{based separation, a promising technology for wastewater treatment \cite{Kumar2019}}. However, \textcolor{black}{the inevitable trade-off between enhancing droplet stability (via size reduction or surfactants) and facilitating demulsification at the separation stage, poses a significant challenge to the large-scale industrial application of ELM technology \cite{Ahmad2011,Kumar2019}.}


The fundamental route to droplet formation is through the fragmentation of a liquid interface subject to shear or impact stresses \cite{Villermaux2007,Barrero2007}. Primary instabilities give rise to corrugated liquid threads or ligaments, whose breakup in turn determines the drop-size distribution. \textcolor{black}{Monodispersity requires precise control over interfacial breakup to suppress uncontrolled local flow instabilities, satellite drop ejection and coalescence. This is routinely achieved via controlled shear-induced emulsification under low-inertia flow conditions by bringing into contact immiscible fluid streams in microfluidic geometries such as Y or T-junctions \cite{Link2004,Garstecki2006,Menech2008}, co-flow \cite{Evangelio2016,Anna2016}, or flow-focusing configurations \cite{Anna2003,Shah2008,Fryd2012,Chen2023}. Recent studies have explored the ability of ``chip-free” microfluidic systems to produce monodisperse droplets via wetting-induced interfacial instabilities and the coupling of impact-induced encapsulation with the Rayleigh–Plateau instability \cite{Visser2018,Su2025}. However, the need for delicate micron-scale channels or nozzles fundamentally restricts scalability, limiting their use in applications.}

\textcolor{black}{In this letter, we report an alternative method which couples horizontal vibration with geometrical confinement to generate vertical standing waves which can act as tuneable sources of monodisperse droplets under horizontal shear. 
We harness an interfacial instability which arises along the end walls of a horizontally vibrated rectangular container filled with two layers of immiscible fluids of sufficiently large density contrast, $\Delta \rho= \rho_\mathrm{l}-\rho_\mathrm{u}$, where l,\,u refer to lower and upper layers, respectively \cite{Talib2007,Jalikop2009}. The horizontal excitation drives a propagating wave that redirects forcing in the vertical direction near the end walls, by harmonically displacing a narrow interfacial front. This front becomes unstable to Faraday waves beyond a critical forcing threshold and the onset of these subharmonic, capillary-gravity waves is captured by a weakly damped Mathieu equation model, where the rate of damping is dominated by dissipation in the Stokes layer near the vertical end-walls \cite{Piao2024}.
The localization of the subharmonic waves at the end walls is regulated by the characteristic decay length of the propagating harmonic wave, $l_\mathrm{d} \sim \sigma/(\nu_\mathrm{m} \Delta \rho \,\omega)$, where $\sigma$ is the surface tension, $\nu_\mathrm{m}=(\rho_\mathrm{u}\nu_\mathrm{u} + \rho_\mathrm{l} \nu_\mathrm{l})/(\rho_\mathrm{u}+\rho_\mathrm{l})$ is the mean kinematic viscosity and $\omega$ the forcing frequency \cite{Bechhoefer1995,Puthenveettil2009}. Hence, the forcing frequency and fluid properties can be used to ensure the regularity of the wave by confining it within a few capillary wavelengths, $l_\mathrm{ca}=[\sigma/(g\Delta \rho)]^{1/2}$, of the end walls. Stable interfacial phenomena are promoted as long as the container's width and depth are at least an order of magnitude larger than the capillary wavelength and the mean Stokes layer thickness, $\delta = \sqrt{2\nu_\mathrm{m}/\omega}$, which prevents the merging of boundary layers   \cite{Lyubimov1986,Talib2007,Sanchez2019}.}


\begin{figure*}
\includegraphics[width=0.8\textwidth]{Revised-Fig.1_2_lower_quality.pdf}
\caption{\label{fig:1} (a) \textcolor{black}{Schematic diagram of the working part of the experiment, with angled side-view images  showing interfacial dynamics near the container’s end walls as a function of increasing forcing acceleration. 
In the angled top-view (orange box),} the pink arrow indicates subharmonic wave tips and the orange arrow points to the edge of the subharmonic Faraday waves right below the onset of droplet formation. 
(b) Top-view image of the formation of regular droplet trains at the end wall, \textcolor{black}{which are separated by half-wavelengths of the Faraday wave, typically of a few millimeters}
(see Movie 1 in Supplemental Materials \cite{SM}). (c) Different vibrational parameters generate droplets of different size, visualized in angled top-view (end wall indicated by yellow line). \textcolor{black}{Several visible satellite droplets were generated before the system reached a stable oscillatory state (see Section S1 in Supplemental Materials \cite{SM}) for details.}}
\end{figure*}

\textcolor{black}{We show that a small increase in vibrational forcing beyond the onset of subharmonic waves can cause subharmonic shedding of nearly identical droplets from the wave tips due to horizontal shear, in contrast to the less regular wave-breaking observed in vertically driven Faraday systems \cite{Puthenveettil2009,Liu2022}. We focus on establishing the conditions required for monodisperse droplet formation and derive universal scalings that reveal the physical mechanisms for droplet detachment and the selection of droplet size. Droplet sources are located every half wavelength which makes the process readily scalable by varying the width of the container, while the forcing frequency regulates the rate of droplet production. Hence, this method demonstrates on-demand generation of tuneable stabilizer-free emulsions sustained by vibration, which rapidly separate upon cessation of forcing, as required by applications such as ELM.}

We performed experiments in a sealed rectangular Perspex container with inner dimensions of $ 170 \times 75 \times 40\, \mathrm {mm}^3$ containing two stably stratified, immiscible liquids of equal volume; see Fig.\ref{fig:1}(a). For the upper layer, we used four grades of silicone oils (polydimethylsiloxane fluid, 10\,cS, 20\,cS, 50\,cS, 100\,cS, Basildon Chemicals Ltd) 
and for the lower layer, two perfluorinated polyethers (Galden\textsuperscript{®} HT135 and HT270,  Solvay), with respective densities, $\rho_\mathrm{u}$, $\rho_\mathrm{l}$, and viscosities, $\nu_\mathrm{u}$, $\nu_\mathrm{l}$ listed in the End Matter. In this letter, we show that monodisperse droplet formation is governed by the \textcolor{black}{kinematic viscosity ratio} $N= \nu_\mathrm{u} /\nu_\mathrm{l}$, and thus we will use this parameter to identify the fluid pairs.

\textcolor{black}{The container was subject to horizontal harmonic oscillations imposed with a horizontal vibration system, with a prescribed velocity $A\omega\cos(\omega t)$, where the amplitude, $A$, and the \textcolor{black}{forcing frequency, $F=\omega/(2 \pi)$}, varied in the ranges $A \le 3.50$\,mm and $20\, \mathrm{Hz} \le F \le 60 \, \mathrm{Hz}$ (see \textcolor{black}{End Matter} for technical specifications)}.
We imaged the interface dynamics in the near-endwall region (orange rectangle in Fig.\ref{fig:1}(a)) with a high-speed camera (Photron FASTCAM mini AX100) and a CMOS camera (Teledyne DALSA  Inc, Genie HC1400). The Photron camera was used to capture time-resolved images in top view and angled top view (Fig.\ref{fig:1}(a)), with respective resolutions of 15.1 and 48.8 pixels/mm and a minimum rate of 4000 frames per second (fps). The droplet formation was monitored by recording top-view movies at 10 fps with the CMOS camera with a maximum resolution of 75.4 pixels/mm.
The interface was visible due to the difference in refractive indices between the two liquid layers. \textcolor{black}{We extracted the wavenumber $k=2\pi/\lambda$ of the interfacial instability, where $\lambda$ is the wavelength of the periodic interface deformation, and the diameter of the droplet $D$ from raw images using image processing, see \textcolor{black}{End Matter} for details}. 

\textcolor{black}{Fig.\ref{fig:1}(a) shows snapshots of the interface between the two fluids near the end wall of the container as the forcing acceleration increases.} The inertial counterflow \textcolor{black}{(red and blue arrows)} induced by the differential acceleration of the fluid layers and the presence of end walls, redirects horizontal oscillatory forcing into vertical oscillation of a localised, \textcolor{black}{flat interfacial front. This front becomes unstable via a Faraday instability at a critical forcing acceleration \cite{Piao2024}.} 
At fixed frequency, a further small amplitude increase beyond the onset of subharmonic waves leads to the onset of droplet detachment at a threshold value $A_c$. \textcolor{black}{Monodisperse droplet formation (Fig.~\ref{fig:1}(c)) is associated with periodic droplet shedding from the crests of the subharmonic waves (see Section S2 in Supplemental Materials \cite{SM} for details), such that each half-wavelength, $\lambda/2$, acts as a subharmonic droplet source, and the number of droplet sources across the container, $2 W/ \lambda$, scales linearly with the width of the container $W$ (Fig.\ref{fig:1}(b)). These sources produce droplets of lower-layer liquid dispersing in the continuous upper layer liquid, which has a relatively large Reynolds number, $\mathrm{Re}=(A\omega)H/\nu_\mathrm{u} = 20-2500$, where $H$=20\,mm is the depth of the fluid layer. The vertical front oscillation near the end walls is accompanied by a mean upper-layer interfacial flow away from the end walls \cite{Gligor2020}. This mean flow enables the transport of droplets from their source to form regular droplet trains.} 
The droplets in each train are of uniform size following short initial transients ($< 10\%$ variation of the mean diameter at $F=35$ Hz, see Fig.\ref{fig:1}(c)), and their size can be adjusted by varying the amplitude and frequency of the applied horizontal forcing.
\begin{figure}[h]
\includegraphics[width=\linewidth]{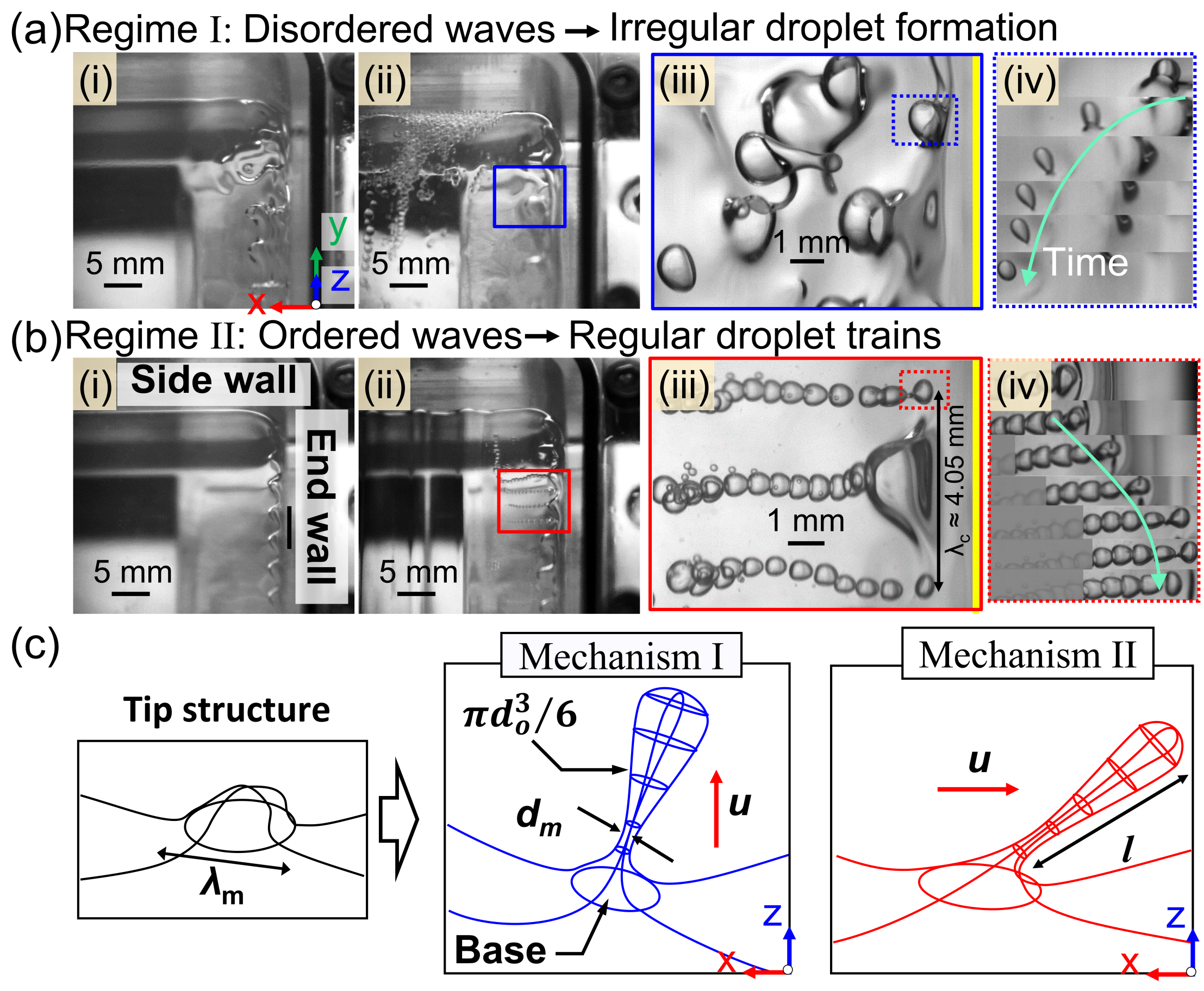}
\caption{\label{fig:2} \textcolor{black}{Distinct regimes of wave and droplet behavior: (a) irregular droplets detaching from disordered waves, (b) regular droplet trains emerging from ordered near-wall waves. Dashed blue and red boxes highlight droplet detachment process. (c) Schematics showing the different types of interfacial breakup observed in (a) and (b). Red arrows indicate the dominant flow direction in the upper layer in the vicinity of the breakup location. }}
\end{figure}

\textcolor{black}{However, monodisperse droplet formation is not the norm, as it depends on the nature of the underlying wave, which varies considerably with vibrational parameters and viscosity \cite{Piao2024}. Fig.\ref{fig:2}(a,b) shows that disordered (ordered) wave patterns near the end walls of the container are associated with irregular (regular) droplet formation. 
We only observe the ordered subharmonic wave patterns required for monodisperse droplet formation (Fig.\ref{fig:2}(b)(i,ii)), when the spread of the harmonic wave away from the end wall, quantified by the damping length scale, is $l_\mathrm{d} \lesssim 2l_{\mathrm{ca}}$ \cite{Piao2024}. Beyond this threshold, subharmonic waves tend to spread further and interact as the forcing acceleration increases, leading to disordered patterns (Fig.\ref{fig:2}(a)(i,ii)). Time-sequences of the detachment of the droplets contained in the highlighted dashed boxes (Fig.\ref{fig:2}(a,b)(iii,iv)), indicate different interfacial breakup mechanisms for ordered and disordered waves, which are distinguished by the orientation along which the wave tip elongates into a ligament. As shown in the schematics of Fig.\ref{fig:2}(c), the wave tips in the irregular case elongate approximately vertically (along $z$, type I breakup) as a result of the subharmonic wave cycle, whereas in the regular case, the larger viscosity ratio strengthens upper-layer shear forces, which act to elongate the ligament in the upper layer in an approximately horizontal direction (along $x$, type II breakup), resulting in approximately monodisperse droplet shedding. }

\begin{figure} [b]
\includegraphics[width=0.75\linewidth]{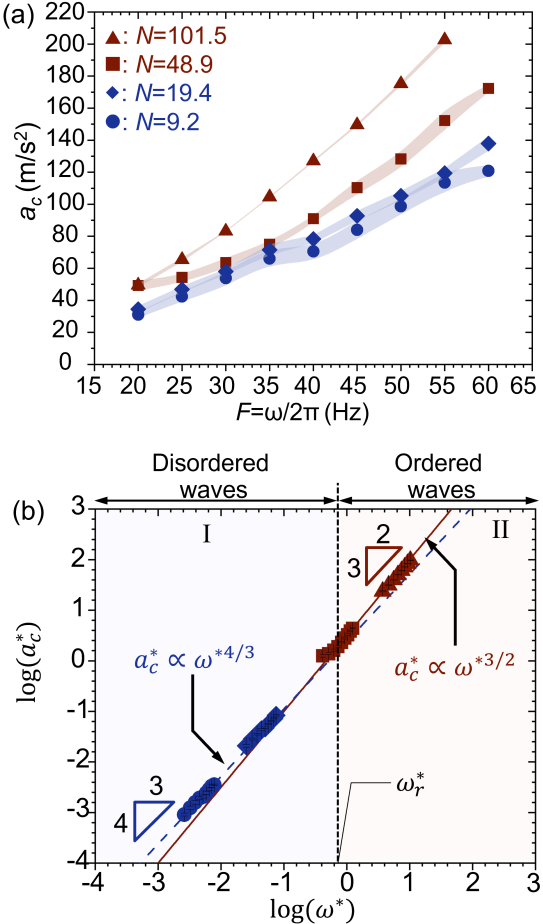}
\caption{\label{fig:3} (a) \textcolor{black}{Variation of the critical acceleration with forcing frequency, $F$, with different symbols distinguishing different viscosity ratios $N$. Red (blue) symbols denotes regular (irregular) droplet formation.} (b) Log-Log plot of non-dimensional acceleration vs non-dimensional frequency, where $\omega_r^*$ marks the transition point between disordered and ordered wave regimes. The linear regressions are shown as a dashed line for the blue data with $\omega^* < \omega_r^*$, and as a solid line for the red data with $\omega^* \geq \omega_r^*$.}
\end{figure}

To gain further insight into these breakup mechanisms, we quantify in Fig.~\ref{fig:3}(a) the threshold acceleration of forcing, $a_\mathrm{c}=A_\mathrm{c}\omega^2$, beyond which droplets appear, as a function of the forcing frequency, for different fluid pairs. 
The lower edge of the shaded bands marks the threshold for droplet formation near the end wall boundaries, while the upper edge indicates the threshold for droplet formation away from these boundaries. We find that the threshold acceleration increases monotonically with increasing forcing frequency for each fluid pair. 
\textcolor{black}{The data is stacked in order of both increasing mean viscosity, $\nu_\mathrm{m}$, and increasing viscosity ratio, $N$. This is because the same lower-layer fluid, HT135, was used in all experiments shown in Fig.~\ref{fig:3}(a). The transition from disordered to ordered waves, and thus, from irregular to regular droplet formation is a function of $\nu_\mathrm{m}$. However, we show our results in terms of viscosity ratio $N$ to highlight that irregular droplet formation thresholds (blue) are insensitive to $N$, whereas regular droplet formation thresholds depend significantly on $N$.}  

\textcolor{black}{In experiments with a $10\times$ more viscous lower layer (HT270) and the lowest value of $N=4.7$, the value of $\nu_\mathrm{m}$ was sufficient to support regular wall-confined subharmonic waves (\textcolor{black}{see Fig.\ref{fig:A2} in the End Matter}), but we did not observe droplet formation.} The waves tips elongated into ligaments oriented horizontally, consistent with type II breakup (Fig.\ref{fig:2}(c)), but not sufficiently to cause rupture. This is because the ratio of the ligament breakup time ($\sim  \lambda \rho_\mathrm{l}\nu_\mathrm{l}/\sigma$) \cite {Eggs2015,Lister1998} to the timescale of \textcolor{black}{the applied viscous force induced by } ($\sim (\omega)^{-1}$) increases in proportion to the lower layer viscosity from $O(10^{-1})$ to $O(10^{0})$, when the fluid HT270 is used as the lower layer instead of HT135.

Given the role of viscous shear in the upper layer on the breakup of the interface, we introduce a viscous-capillary length, $l_\mathrm{vc}=\nu_\mathrm{u}^{2}/(\sigma/\rho_\mathrm{u})$, and a viscous-capillary frequency, $\Omega_\mathrm{vc}=(\sigma/\rho_\mathrm{u})^2/(\nu_\mathrm{u}^3)$ \cite{Goodridge1997, Piao2024}, which we use to scale the critical acceleration, $a_\mathrm{c}^{*}=a_\mathrm{c}/(l_\mathrm{vc} \Omega_\mathrm{vc}^{2})$, and frequency, $\omega^{*}=\omega/ \Omega_\mathrm{vc}$. 
\textcolor{black}{The scaled data in Fig.\ref{fig:3}(b) aligns in order of increasing upper layer viscosity along the non-dimensional frequency axis. 
We pinpoint the threshold frequency between disordered and ordered wave regimes, $\omega_\mathrm{r}^* \approx 0.72$, which is associated with the non-dimensional damping length, $l_{\mathrm{d}}/l_{\mathrm{ca}}\approx 2$. Based on this criterion, we re-examine the lower frequency range of the data at $N=48.5$, and find that the breakup mechanism is visibly evolving away from type II (horizontal elongation) towards type I (vertical elongation) for values right below $\omega_r^*$.}

To understand the physics governing both irregular and regular interfacial breakup, we develop two models of the elongated fluid ligament based on a simple force balance (type I), and an energy balance (type II), using suitable scaling arguments \cite{Goodridge1997}.
The model for type I breakup is based on the assumption of capillary instability \cite{Eggs2008}: as the wave tip with the most unstable wavelength $\lambda_{\mathrm{m}}$ elongates into a thin ligament with a minimum diameter $d_m$ due to \textcolor{black}{vertical shear forces} in the upper fluid layer, the ligament can rupture when the viscous shear force overcomes the capillary restoring force. \textcolor{black}{The capillary force per unit mass on the droplet-forming ligament with a mean diameter $d_o$ is  $F_{\sigma}=\sigma \pi d_o/(\pi \rho_\mathrm{l} d_o^3/6)$. The mean diameter is approximately proportional to the \textcolor{black}{characteristic length scale associated with the wave tip}, $d_o \sim \lambda_\mathrm{m}$, which is approximately the onset wavelength ($\lambda_\mathrm{c}$) at droplet formation \cite{Lang1962,Goodridge1997, Marmottant2004, Piao2021}. Hence, $F_{\sigma} \sim  \sigma /(\rho_\mathrm{l} \lambda_\mathrm{c}^2)$,  where $\lambda_\mathrm{c}=2\pi/k_\mathrm{c}$ in the capillary limit is given by $\omega^2=4\sigma (k_\mathrm{c})^3/(\rho_\mathrm{l}+\rho_\mathrm{u})$ \cite{Piao2024}. The vertical shear force exerted by the fluid in the upper layer in the end wall region is estimated using the
vibrational velocity $A\omega$ and the thickness of the upper Stokes layer \textcolor{black}{$\delta_\mathrm{u}$} such that $F_{V} = \nu_\mathrm{u} \nabla^2 u \sim \nu_\mathrm{u} (A\omega) / (\delta_\mathrm{u})^2 \sim  A\omega^2$. The threshold for droplet detachment is $F_{V} \sim F_{\sigma}$ at $A=A_\mathrm{c}$.} Hence, the critical acceleration on the viscous-capillary scale for the onset of \textcolor{black}{type I} breakup is given by
\begin{equation}
a_\mathrm{c}^{*} \sim  (\rho_\mathrm{u}/\rho_\mathrm{l})^{1/3} {\left(\frac{\rho_\mathrm{l}+\rho_\mathrm{u}}{\rho_\mathrm{l}}\right)}^{2/3} \omega^{*4/3}.
\label{equ1}
\end{equation}
\textcolor{black}{Fitting of the blue experimental data in Fig.~\ref{fig:3}(b) as a function of the right handside of Eq.\,(\ref{equ1}), albeit spanning a limited range of $\omega^*$, yields a proportional relationship which supports the scaling in Eq.\,(\ref{equ1}), see Fig.\ref{fig:A3}(a) in the End Matter. }

For type II breakup (Fig.\ref{fig:2}(c)), we argue that the energy received to create regular droplets ($E_\mathrm{s}$) is set by the balance between the work of the horizontal shear force exerted by the upper-layer fluid ($E$) and the energy dissipation due to the resistance to extension of the droplet-forming ligament. For fixed forcing, the constant rate of droplet production indicates that the rate of energy transfer to the droplet-forming ligament $\dot{E_\mathrm{s}}$ is zero. Hence, the onset of regular droplet formation can be predicted by balancing the rate of work of the horizontal shear force and the rate of energy dissipation. \textcolor{black}{The rate of work per unit mass can be written as $\dot{E}=F_H A\omega$, where $F_H\sim A\omega^2$ is the horizontal viscous shear force per unit mass which takes a form similar to $F_V$ above. The rate of energy dissipation per unit mass is generally estimated as $\varepsilon=\nu_\mathrm{l} (u_\mathrm{s}/l)^2$, where $\nu_\mathrm{l}$ is the kinematic viscosity of the droplet-forming ligament, $u_\mathrm{s}$ and $l$ represent the stretching velocity and length of the ligament, respectively \cite{Goodridge1997,Piao2021}. Using $\omega^{-1}$ as the time scale of the stretched ligament, the stretching velocity is $u_\mathrm{s}=l \omega$ and, thus, the energy dissipation rate is given by $\varepsilon \sim \nu_\mathrm{l} \omega^2$ \cite{Goodridge1997,Piao2021}.
Balancing $\dot{E}$ and $\varepsilon$ on the viscous-capillary scale yields an expression for the critical acceleration,
}
 \begin{equation}
a_\mathrm{c}^{*} \sim (\nu_\mathrm{l}/\nu_\mathrm{u})^{1/2} \omega^{*3/2}=N^{-1/2} \omega^{*3/2}.
\label{equ2}
\end{equation}
\textcolor{black}{Eq.\,(\ref{equ2}) indicates that the threshold acceleration is governed by the viscosity ratio, rather than the viscosity of the upper fluid layer or the mean viscosity that governs the onset of subharmonic waves. This is also the case in the experiment, as shown in Fig.\ref{fig:3}(a). A linear fit of the red experimental data in Fig.~\ref{fig:3}(b) as a function of the right handside of Eq.\,(\ref{equ2}) also yields an approximate proportional relation which supports the scaling in Eq.\,(\ref{equ2}) (see Fig.\ref{fig:A3}(b) in the End Matter)}.

\begin{figure}[t]
\includegraphics[width=0.9\linewidth]{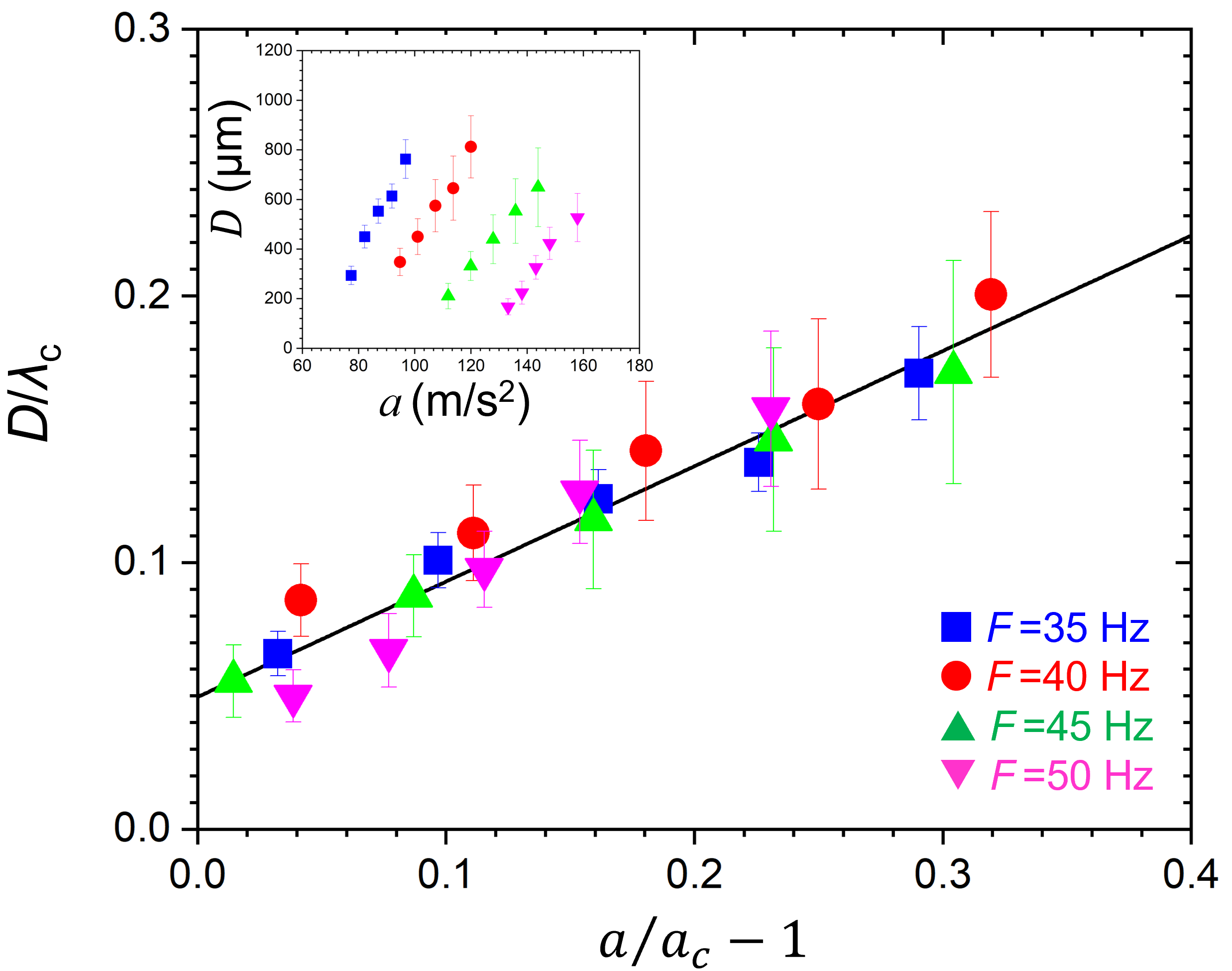}
\caption{\label{fig:4} Droplet diameter normalised by $\lambda_\mathrm{c}$ as a function of non-dimensional forcing acceleration $a/a_\mathrm{c}$ for regular droplet formation at $N$=48.9 across different forcing frequencies $F$. The solid line is a linear fit to the data. Inset: dimensional droplet diameter vs forcing acceleration. Error bars denote the standard deviations of multiple measurements. }
\end{figure}

\textcolor{black}{We now turn to the selection of the droplet diameter, $D$, which increases linearly with forcing acceleration, see inset of Fig.~\ref{fig:4}. We find that the ratio of the standard deviation of the droplet diameter to the mean diameter is smaller than 25\% for all tested conditions,
which meets the accepted criterion for monodisperse emulsions \cite{Ho2022}. The minimum measured droplet diameter, $D=167.0 \pm 32.7$\,µm, was reached for the largest frequency, $F=50$\,Hz. Fig.~\ref{fig:4} shows that the scaled data, $D/\lambda_\mathrm{c}$, falls on a single straight line within experimental uncertainty, when plotted against the excess acceleration from onset. This is because $\lambda_\mathrm{c}$ varies approximately as $F^{-2/3}$ and substituting the scaling relation for $a_c$ from Eq.\,(\ref{equ2}), we get $a/a_\mathrm{c} \sim A/\delta_\mathrm{l}$, which has a $F^{1/2}$ dependence. This leaves a weak frequency dependence in $F^{-1/6}$ which we cannot resolve experimentally over the frequency range investigated. To within experimental resolution, the scaled droplet size at onset is $D/\lambda_{\mathrm{c}} \simeq 0.05$. This result is consistent with Lang’s empirical model obtained for ultrasonic atomization and emulsification \cite{Li1978,Li2014}. The value is small because the droplet size is set by the size of the wave tip rather than the actual wavelength at onset, but  $\lambda_\mathrm{m} \sim \lambda_\mathrm{c}$, as suggested by the angled top-view image in Fig.~\ref{fig:1}(a).}  
\textcolor{black}{Extrapolating from our measurements at 50\,Hz, we estimate that frequencies on the order of 3\,kHz would be required to yield 10\,µm droplets. Despite producing larger droplets than ultrasonic emulsification ($\sim 1$\,µm), our method which applies in the sub-ultrasonic regime, offers enhanced process control with the dual advantages of monodisperse emulsification and tuneable droplet sizes. This is because the scaled droplet size is simply set by the ratio of the forcing amplitude to the thickness of the interfacial Stokes boundary layer in the lower layer, which regulates wave tip elongation. Moreover, the number of droplet sources scales with container width, enabling spatially-extended emulsification in contrast to the localized nature of ultrasonic emulsification because of the rapid decay of the acoustic energy near the probe tip.}

In summary, we have studied the formation of a monodisperse emulsion at the interface between two confined layers of immiscible liquids subject to horizontal vibrations. We have shown that monodisperse micro-scale emulsions \textcolor {black}{are generated for sufficiently large viscosity ratios and forcing frequencies to support ordered subharmonic waves localised near the end walls. 
We demonstrate a novel emulsification method that couples vertical oscillations (to induce Faraday waves) with horizontal shear, which facilitates droplet detachment and transport away from the near-wall region (see Section S3 in Supplemental
Materials \cite{SM} for further discussion.)} The simple dependence of the non-dimensional droplet diameter on the excess acceleration beyond onset indicates that the diameter of monodisperse droplets can be easily controlled by adjusting the forcing. A key advantage of our method is that clogging, which is common in microfluidics \cite{Shah2008,Quintero2018}, can be avoided during the generation of monodisperse droplets. \textcolor{black}{Droplet generators occur every half wavelength ($\lambda/2$) along the chamber width ($W$), such that scale-up production is achieved by widening the chamber to accommodate $2W/\lambda$ sources operating at frequency $F/2$}. \textcolor{black}{To broaden the applicability of the proposed emulsification mechanism, ongoing work is investigating the influence of surfactants on droplet pinch-off dynamics. Future studies will also explore alternative configurations that enable the coupling of vertical and horizontal interfacial vibration, such as vibrating plates \cite{Xu2021} or pulse-driven two-fluid systems.}

{\em Acknowledgments} -- This research was supported by a Horizon Europe Guarantee MSCA Postdoctoral Fellowship through EPSRC (EP/X023176/1).

{\em Data availability} -- The data that support the findings of this article are openly available \footnote{See Data\_Piao.gz in the Supplemental Material.}.

\bibliography{revision_final}
\clearpage


\appendix
\renewcommand{\thefigure}{A\arabic{figure}}  
\setcounter{figure}{0}  

\section{End Matter}

\subsection{Horizontal vibration system and fluid properties}
\begin{figure}[h]
\centering
\includegraphics[width=1\linewidth]{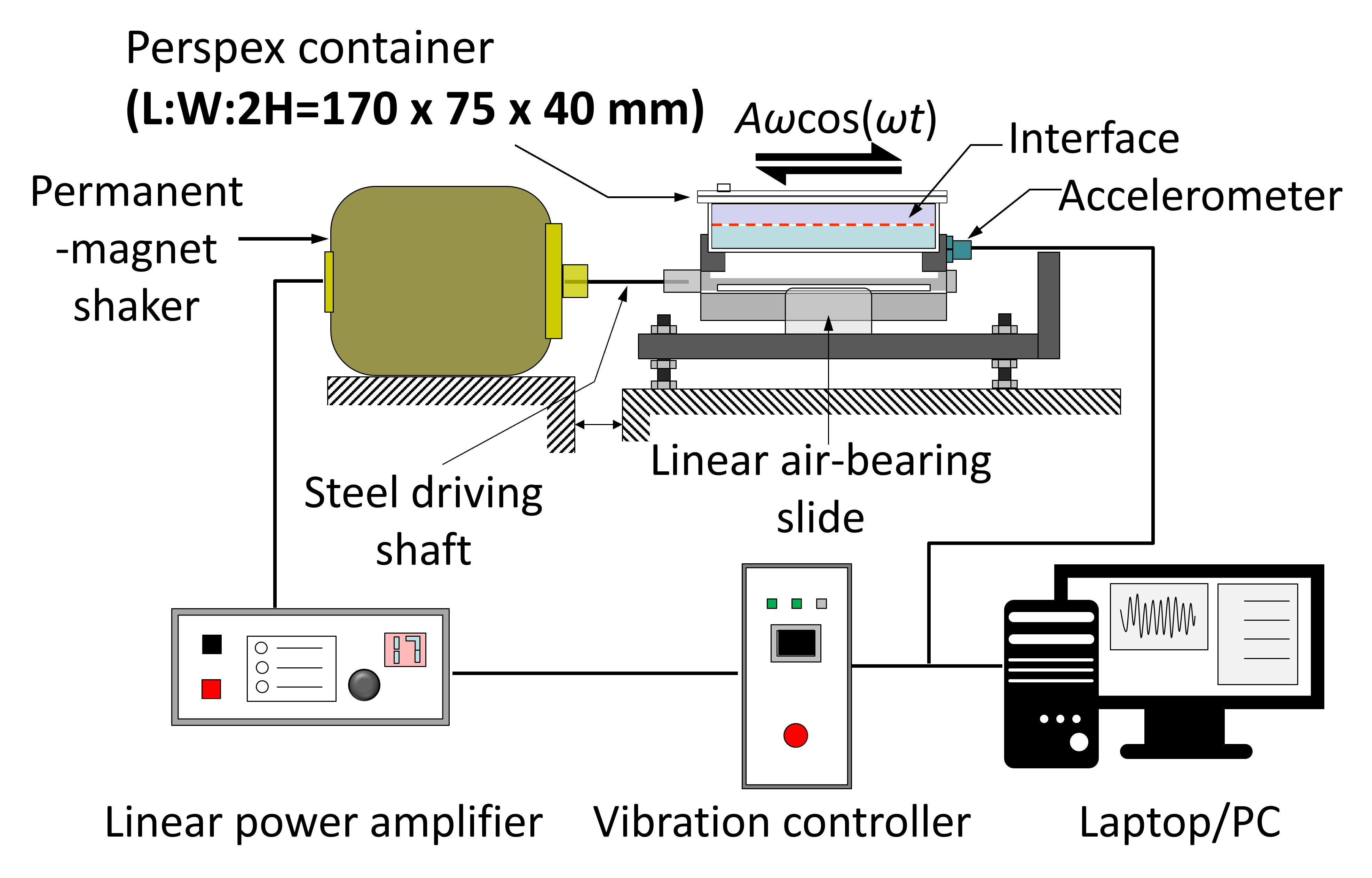}
\caption{Schematic diagram of the horizontal vibration system.}
\label{fig:A1}
\end{figure}

Fig.~\ref{fig:A1} shows a schematic diagram of the experimental setup. The Perspex container is rigidly mounted on a linear air-bearing slide, which is connected to a permanent magnet shaker  (LDS-V450) via a steel driving shaft (a thin rod) with a diameter of 2 mm. A vibration controller (LDS-COMET USB), interfaced to a laptop or PC, is used in combination with a linear amplifier (LDS, PA1000L) to control the shaker, which imposes horizontal vibration on the container with a prescribed velocity. The harmonic content of the container's motion, measured using a calibrated accelerometer (PCB Piezotronics, model 353B43), remained below 0.1\% across the entire frequency range. This vibration system has previously been employed to investigate wave instabilities \cite{Talib2007,Jalikop2009, Piao2024}. 

\textcolor{black}{The density and viscosity of the fluids are listed in table \ref{tab:1}.}

\begin{table}[h]
  \begin{center}
\def~{\hphantom{0}}
  \begin{tabular}{lcc}
           Upper layer& $v_\mathrm{u} (10^{-6}$ m$^2$/s)&$\rho_\mathrm{u}$ (kg/m$^3$)  
\\[3pt]
          Silicone oil (10 cS)& 10.3&935 \\
          Silicone oil (20 cS)& 21.7&950 \\ 
          Silicone oil (50 cS)& 54.8&961  \\
          Silicone oil (100 cS)& 113.7&961 
\\[3pt]
Lower layer& $v_\mathrm{l} (10^{-6}$ m$^2$/s)&$\rho_\mathrm{l}$ (kg/m$^3$) 
\\[3pt]
Galden HT 135& 1.12$^b$&1752  \\ 
Galden HT 270& 11.7$^a$&1856$^a$ \\
  \end{tabular}
  \caption{Physical properties of the liquids used in the experiments. The viscosities of silicone oils were measured at $21\pm1$°C using a Kinexus rheometer. $^a$From manufacturer's data at 25°C. $^b$Measured by \cite{Jalikop2009} at 21$\pm1$°C.}
  \label{tab:1}
  \end{center}
\end{table} 
\begin{figure}[b]
\includegraphics[width=0.75\columnwidth]{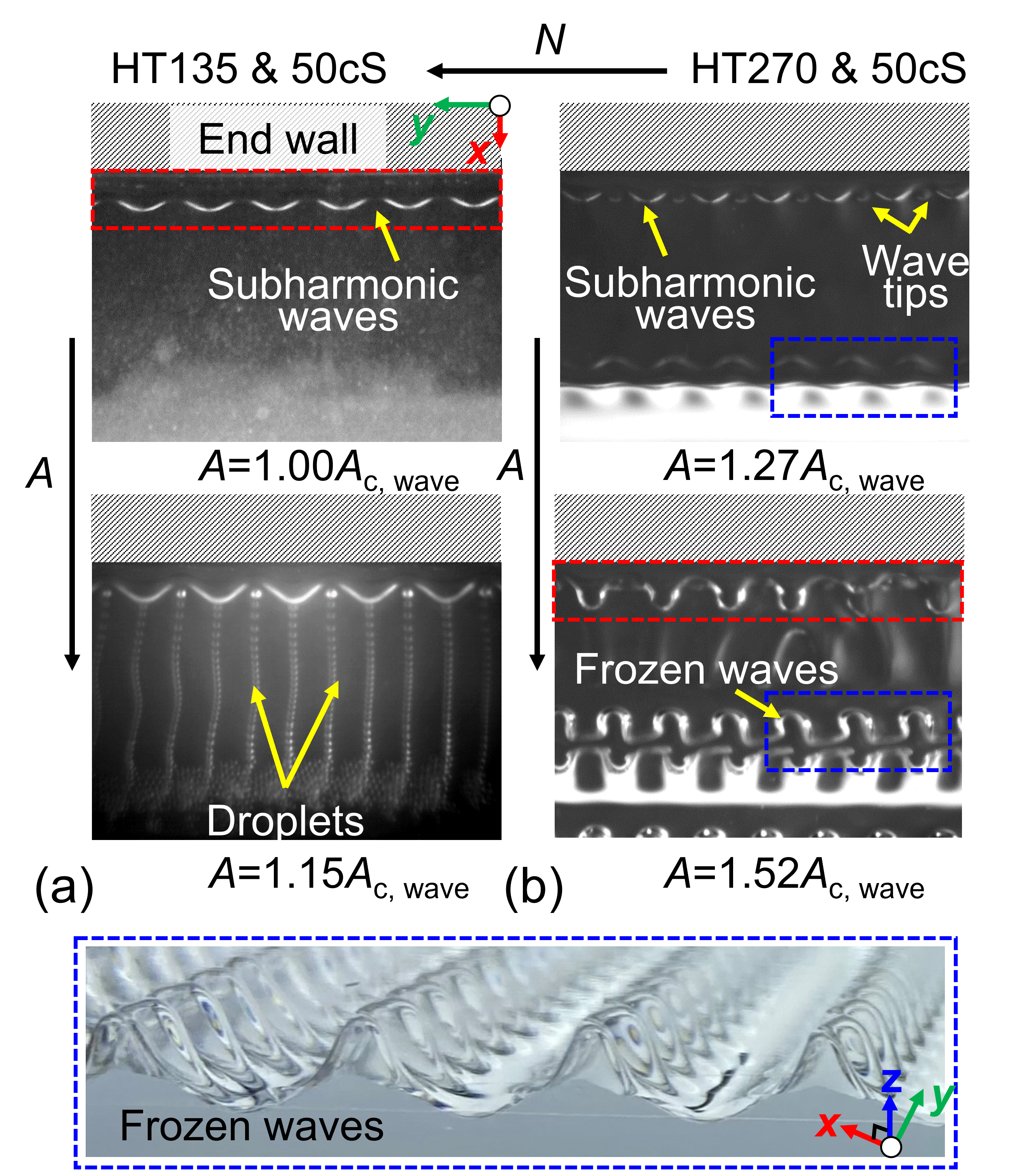}
\caption{\label{fig:A2} \textcolor{black}{Interfacial instabilities near the end wall for fluid pairs:  (a) $N=48.9$ (HT135 \& 50cS) and (b) $N=4.7$ (HT270 \& 50cS) at a forcing frequency of $F=40$ Hz. Images are captured from the top view and the red box indicates the region near the end wall. The coordinate system shown was defined as in Fig. \ref{fig:1}(a) of the main text.  Frozen waves are highlighted by blue boxes in (b), while the bottom-inserted blue box provides a representative side view.}}
\end{figure}
\subsection{Measurement of wavenumber, droplet diameter, and threshold acceleration}
To obtain precise wavenumber information, a pattern with 12 black lines parallel to the end walls was placed below the transparent bottom boundary of the container to assist measurements. The lines have a thickness of 0.2 mm and an interline spacing of 1 mm. The visualisation in top-view of the image of this line pattern, refracted by the fluid interface, enabled the detection of small interface deformations along the $y$-direction (Fig.\ref{fig:1}(a))) down to 0.1 mm, by measring the distortion of the line. The wavenumber of the periodic deformation, $k=2\pi/\lambda=2\pi M/(W/2)$ \cite{Piao2024}, was determined by measuring $M$, the number of wavelengths ($\lambda$) spanning the half-width ($W/2$) of the container. $M$ was obtained by Fast Fourier Transform of the edge of the wave pattern along the $y$-direction tracked using MATLAB's Canny algorithm. 
\begin{figure*}[t]
\includegraphics[width=0.75\linewidth]{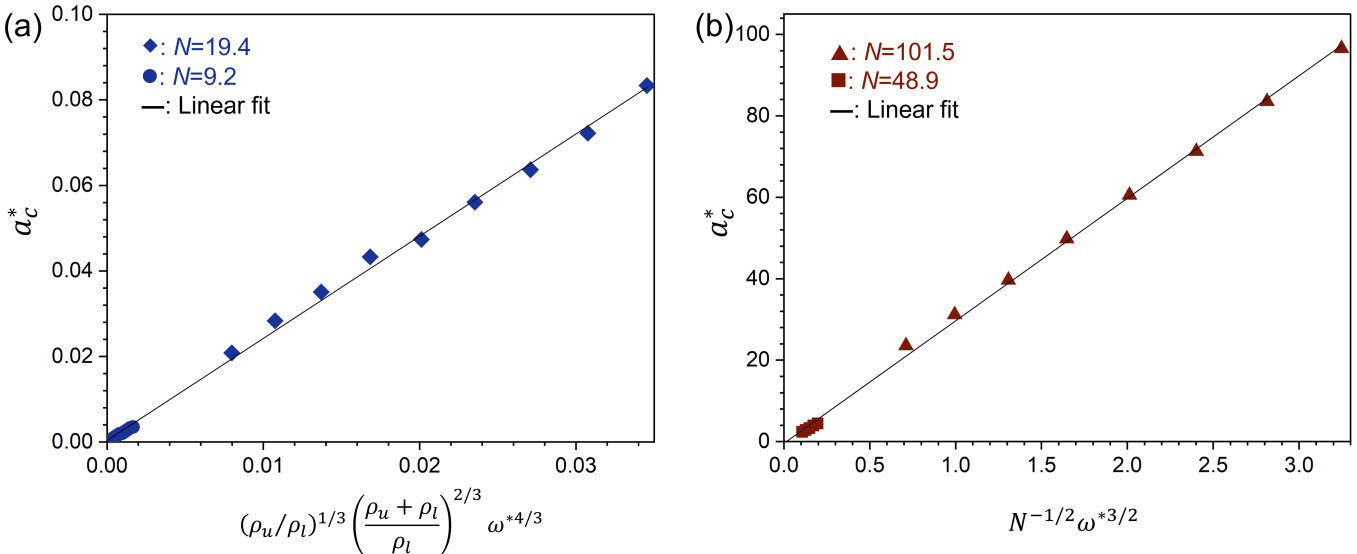}
\caption{\label{fig:A3} \textcolor{black}{(a) Comparison between the experimental data and the linear fit of the model (\ref{equ1}) for $N=9.2$ and $N=19.4$, yielding a coefficient of determination $R^2=0.99801$. (b) Comparison between the experimental data and the linear fit of the model (\ref{equ2}) for $N=48.9$ and $N=101.5$ in the ordered wave regime, showing good agreement ($R^2=0.99895$).}}
\end{figure*}

The boundaries of dispersed droplets obtained from our experimental images exhibit strong contrast and clear definition. Therefore, we extracted the edge of the droplet and calculated its diameter $ D=\frac{2}{J}\Sigma_{i=1}^{J} \sqrt{(x_i-x_\mathrm{c})^2+(y_i-y_\mathrm{c})^2}$, where $J$ is the number of edge points, ($x_i,y_i$) the coordinates of an edge point and ($x_\mathrm{c}, y_\mathrm{c}$) the centre of the droplet.

\textcolor{black}{The threshold acceleration measurements were made by imposing forcing frequency and gradually increasing the forcing amplitude in increments of 0.01\,mm until the first appearance of droplets at the threshold amplitude $A_c$. We recorded at least 750 oscillation cycles to monitor the collection of droplets at onset (see Movie 2 in Supplemental Materials \cite{SM}) and ensured that all droplets coalesced back into the lower layer before taking the next measurement}.
\vspace{-5mm}
\subsection{\textcolor{black}{Absence of near-wall droplet detachment for $N< 48.9$ under ordered wave conditions}}
\textcolor{black}{Subharmonic waves excited near the end walls vary according to the relative influence of viscous and capillary effects, which in our experiments are imposed by vibrational parameters and the mean viscosity of the fluid \cite{Piao2024}. However, as noted in the main text, the droplet detachment\textemdash closely related to the properties of these waves\textemdash is strongly governed by the viscosity ratio. In Fig.\ref{fig:A2}, we present different interfacial phenomena near the end wall with the fluid pairs HT135 $\&$ 50cS (Fig.\ref{fig:A2}(a)) and HT270 $\&$ 50cS (Fig.\ref{fig:A2}(b)). All experiments were performed with $F =$ 40 Hz and in each figure the forcing amplitudes increased from top to bottom. In both fluid pairs, regular subharmonic waves emerged near the end wall when the amplitude $A$ exceeded the critical value for their onset, $A_\mathrm{c,wave}$. For the HT135 $\&$ 50cS fluid pair ($N=48.9$), increasing forcing amplitude causes droplet detachment from subharmonic wave tips. In contrast, for the HT270 $\&$ 50cS fluid pair ($N=4.7$), wave tip breakup is absent; instead, other instabilities, such as frozen waves \cite{Jalikop2009}, spread toward the end wall.}
\vspace{-3mm}
\subsection{\textcolor{black}{Comparison between scaling laws and experimental data}}
\textcolor{black}{We quantify the accuracy of the scaling relations obtained in equations (\ref{equ1}) and (\ref{equ2}) in the main text by plotting the critical acceleration measured experimentally as a function of its scaling relation in Fig.~\ref{fig:A3}(a) and Fig.~\ref{fig:A3}(b), respectively. Each set of experimental data contains measurements for two values of viscosity ratio and a range of frequencies.} 
\textcolor{black}{In Fig.~\ref{fig:A3}(a), 
a linear fit to the data with a coefficient of determination, $R^2 = 0.99801$, yields a negligible intercept, $\mathrm{a}=3.27 \times 10^{-4}$. Thus, $a^*_c$ is effectively proportional to $(\rho_\mathrm{u}/\rho_\mathrm{l})^{1/3} {(\frac{\rho_\mathrm{l}+\rho_\mathrm{u}}{\rho_\mathrm{l}}})^{2/3} \omega^{*4/3}$ with a slope of $2.39\pm0.03$. In Fig.~\ref{fig:A3}(b), the linear fit has a coefficient of determination, $R^2 =0.99895$. The intercept, $-0.46\pm 0.44$, is small and vanishes approximately to within error bars. Hence, the critical acceleration $a^*_c$ is effectively proportional to $N^{-1/2} \omega^{*3/2}$ with a slope of $30.10\pm0.28$.}

\end{document}